\documentclass[final]{aipproc}
\layoutstyle{6x9}

 \newcommand\beq{\begin{equation}}
 \newcommand\eeq{\end{equation}}
 \newcommand\beqn{\begin{eqnarray}}
 \newcommand\eeqn{\end{eqnarray}}
\def\Im{{\rm Im\,}}
\def\Re{{\rm Re\,}}
\def\la{\langle} 
\def\ra{\rangle} 
\def\mb{\,\mbox{mb}}
\def\fm{\,\mbox{fm}} 
\def\GeV{\,\mbox{GeV}}

\def\Pom{{I\!\!P}}

\begin{document}

\title{Small Angle Scattering of Polarized Protons}

\author{B.Z.~Kopeliovich}{address={Max-Planck Institut f\"ur Kernphysik, 
Postfach 103980, 69029 Heidelberg\\ Institut f\"ur Theoretische Physik 
der Universit\"at, D-93040 Regensburg,\\ Joint Institute for Nuclear 
Research, Dubna, Russia}}

\begin{abstract}

Experiment E950 at AGS, BNL has provided data with high statistics for
the left-right asymmetry of proton-carbon elastic scattering in the
Coulomb-nuclear interference region of momentum transfer. It allows to
access information about spin properties of the Pomeron and has practical
implications for polarimetry at high energies. Relying on Regge
factorization the results for the parameter $r_5$, ratio of spin-flip to
non-flip amplitudes, is compared with the same parameter measured earlier
in pion-proton elastic and charge exchange scattering. While data for
Im$r_5$ agree (within large systematic errors), there might be a problem
for Re$r_5$. The $\pi N$ data indicate at a rather small contribution of
the f-Reggeon to the spin-flip part of the iso-scalar amplitude which is
dominated by the Pomeron.  This conclusion is supported by direct
analysis of data for elastic and charge exchange $pp$ and $pn$ scattering
which also indicate at a vanishing real part of the hadronic spin-flip
amplitude at energies $20\GeV$ and higher. This is a good news for
polarimetry, since the E950 results enhanced by forthcoming new
measurements at AGS can be safely used for polarimetry at RHIC at higher
energies.
 \end{abstract}

\maketitle

\section{Introduction}

It is usually assumed that small angle elastic scattering at high
energies is dominated by Pomeron exchange. At the same time, 
definition for the Pomeron varies depending on a model (a Regge pole,
pole plus cuts, two-gluon model, DGLAP, BFKL, two-component Pomeron, 
etc.) what
led to a confusion among the community. In what follows, we do not assume
any model, unless otherwise specified. We treat the Pomeron as a shadow
of inelastic processes. i.e. the dominant contribution to the elastic
amplitude which has vacuum quantum numbers in the crossed channel and is
related to the main bulk of inelastic channels via the unitarity 
relation.

Here we are interested in spin properties of such a shadow, namely, the
spin-flip part of the elastic $pp$ amplitude related to the Pomeron.
Naively, treating the Pomeron perturbatively, one may expect it to
conserve $s$-channel helicity as the quark-gluon vertex does. However,
even perturbatively a quark gains a substantial anomalous color-magnetic
moment which leads to a spin-flip, like it happens in QED in $g-2$
experiments. Besides, there are many nonperturbative mechanisms
generating a Pomeron spin-flip, which are overviewed in \cite{review}.

We will present the results in terms of the parameter $r_5$
which is defined in \cite{review} and is proportional to ratio of the 
spin-flip to non-flip forward elastic amplitudes. 
 \beq
r_5 = \frac{2m_N\Phi_5}
{\sqrt{-t}\,\Im(\Phi_1+\Phi_3)}\ ,
\label{3}
 \eeq
 where the helicity amplitudes are defined as,
 \beq
\Phi_1 = \la ++|\hat M|++\ra\ ;\ \ \ 
\Phi_3 = \la +-|\hat M|+-\ra\ ;\ \ \ 
\Phi_5 = \la ++|\hat M|+-\ra\ .
\label{6}
 \eeq
 Parameter $r_5$ may vary with energy, in particular, it is expected to
rise \cite{kp}.

In this talk I focus on the best of our knowledge of $r_5(s)$ which is
still a challenge. Importance of this task is two-fold. First of all, it 
reflects the underlying dynamics and data for $r_5$ should be compared 
with numerous and diverse model predictions. Second of all, the 
polarization program at RHIC needs reliable and fast polarimetry.
The currently available polarimeter, is based on the 
effect of Coulomb-nuclear interference (CNI) \cite{kl,bgl} which is fully 
predicted theoretically provided that $r_5$ is known.

\section{CNI revisited}

It is not easy to access the spin-flip part of the Pomeron amplitude
since it hardly contributes to single spin asymmetry $A_N(t)$. Indeed,
although the Pomeron is not a Regge pole, but if $r_5$ does not vary
steeply with energy, one should not expect a large phase shift between
the spin-flip and non-flip parts of the Pomeron amplitude. Of course
$r_5$ can be extracted from spin correlation $A_{SL}$ which is, however,
difficult to measure.

A unique source of a spin-flip amplitude with a right phase (i.e. with
about $90^0$ phase shift) is Coulomb scattering. This real amplitude
proportional to the anomalous magnetic momentum of proton, interferes with
the imaginary non-flip part of the Pomeron amplitude leading to a
sizeable spin asymmetry $A_N$ which is nearly independent of energy. The
latter fact, as well as possibility to predict the effect, are crucial
for polarimetry at high energy. 

If the spin-flip part of the Pomeron amplitude were zero, the CNI 
contribution to single spin asymmetry would be fully predicted \cite{kl},
 \beq
A_N(t) = \frac{4\,(t/t_p)^{3/2}}
{3(t/t_p)^2 + 1}\ A_N(t_p)\ .
\label{10}
 \eeq
 Here
 \beq
t_p = - 8\sqrt{3}\,\frac{\pi\,\alpha}{\sigma^{pp}_{tot}}\ ,
\label{20}
 \eeq
 is the position of the maximum of $A_N(t)$ which is equal to
 \beq
A_N(t_p) = \frac{\sqrt{-3t_p}}{4m_N}\,(\mu_p-1)\ ,
\label{30}
 \eeq
 where $\mu_p-1\approx 1.79$ is the anomalous magnetic moment of the 
proton.

\begin{figure} 
\includegraphics[height=.3\textheight]{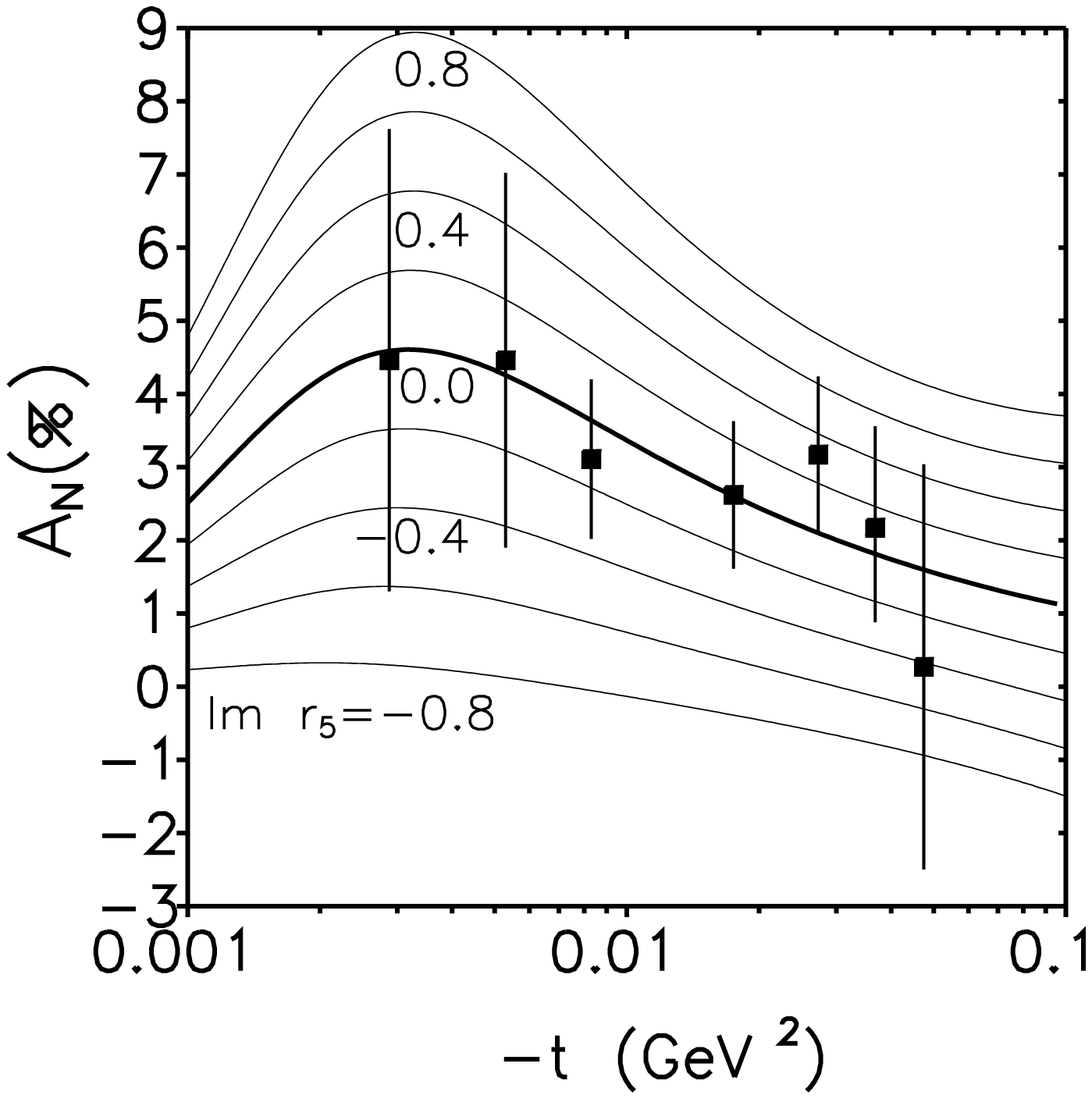}
\includegraphics[height=.3\textheight]{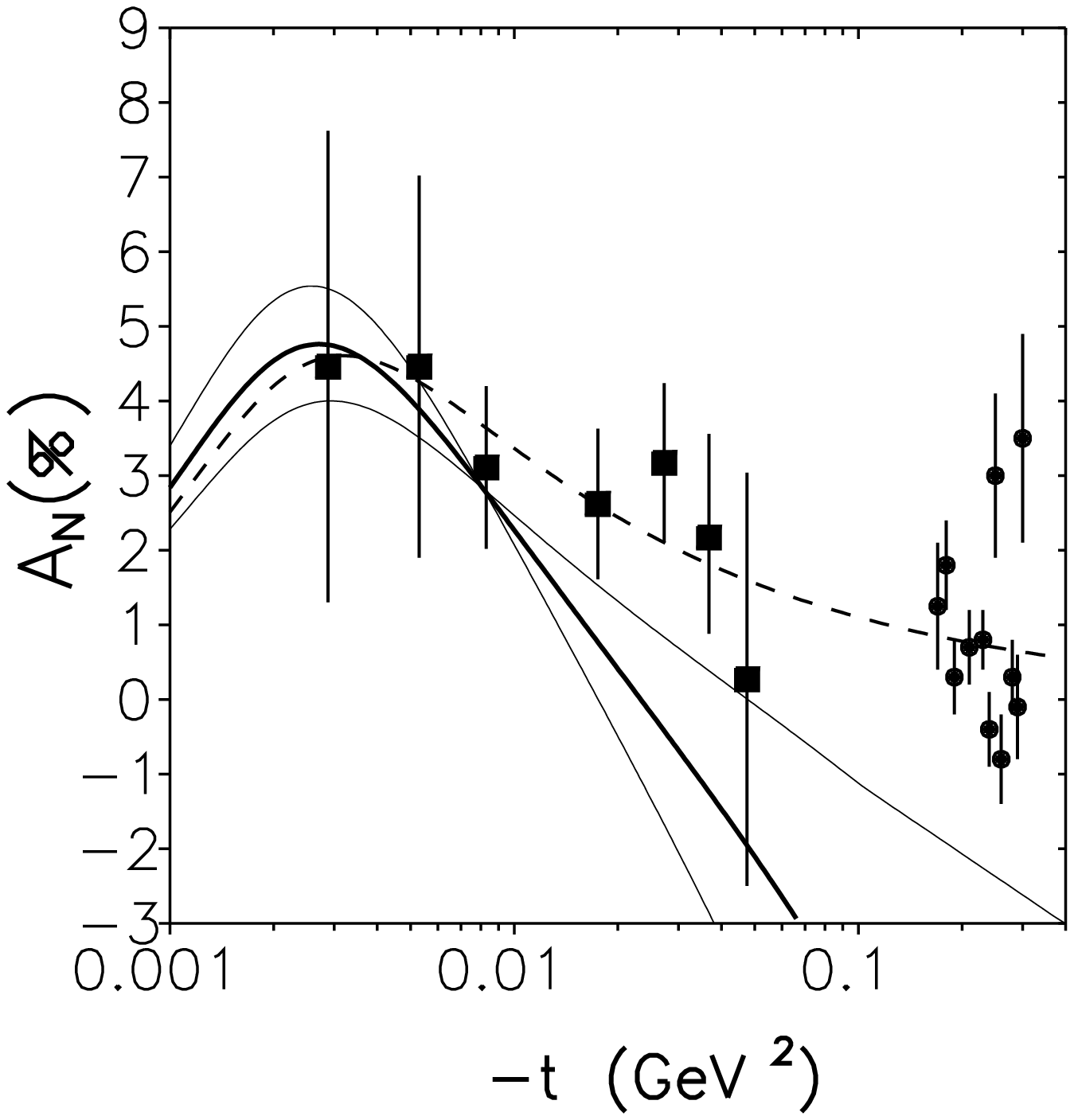} 
 \caption{Data from the E704 experiment at Fermilab \cite{e704} (squares)
compared to theoretical calculations. Left panel: calculations with
$\rho=0$, $\delta=0$ and $\Re r_5=0$ for different values of $-0.8 < \Im
r_5 < 0.8$. Thick curve corresponds to $r_5=0$.  Right panel: $r_5$
correspond to the results of the E950 experiment at BNL \cite{e950} as
given by (\ref{70}). Thin curves show the corridor of uncertainty. Round
points show results of other experiments, see in \cite{e704}. The dashed
curve corresponds to $r_5=0$.}
 \label{pp-pomm}
 \end{figure}

Predictions for $A_N(t)$ \cite{kl} shown by thick curve in 
Fig.~\ref{pp-pomm} (left panel) are compared with data from the 
experiment
E704 at $200\GeV$ \cite{e704}. Apparently, agreement is good, although 
the error bars are quite large.
 
One may conclude that CNI provides a perfect absolute polarimeter which
can be safely used at high energies. Life, however, is more difficult,
but also more exciting. The Pomeron amplitude may have a nonzero
spin-flip part $r_5$ which affects the spin asymmetry \cite{kz,larry}.  
Fig.~\ref{pp-pomm} (left panel) demonstrates how $A_N(t)$ varies versus
$\Im r_5$ assuming $\Re r_5=0$ .

Such a sensitivity to $r_5$ of the CNI effects leads to two-fold
consequences:
 \begin{itemize}
 \item CNI polarimetry turns out to be less certain than has been
originally expected;
 \item $A_N$ in the CNI region is an observable maximally sensitive to
$r_5$ and can be used to determine the magnitude of the Pomeron
spin-flip.
 \end{itemize}
  
\section{CNI with nuclear targets: the E950 experiment}

In order to make use of a CNI polarimeter one should first of all
calibrate it, i.e. perform measurements of $r_5$ with proton beams or
target with known polarization. Such beams are available at AGS, but only
at energies not much above $20\GeV$.  In this energy range contribution
of the sub-leading Reggeons is still important and can substantially
contribute to $r_5$ giving it a steep energy dependence. It would be too
risky to rely on a value of $r_5$ measured at these energies for
polarimetry at much higher energies. Especially dangerous are the
iso-vector Reggens $\rho$ and $a_2$ which are spin-flip dominated. To get
rid of these unwanted contributions it was suggested in \cite{mine} to
use CNI on iso-scalar nuclei, in particular carbon. However, two
important questions were raised:
 \begin{itemize}
 \item Can one use $r_5$ measure on nuclear target for polarimetry in 
$pp$ 
scattering?
\item How should expression for CNI asymmetry be modified in the case of 
nuclear target?
\end{itemize}
 As for the first question, it has been known since 50s \cite{bethe}
that $r_5$ remains unchanged, if to treat nuclear effects within
the optical model. An updated proof and discussion of possible
corrections can be found in \cite{kt}, as well as the expression for 
$r_5$ on a nuclear target,
 \beq
r^{pA}_5(t) = \frac{1-i\rho_{pA}(t)}
{1-i\rho_{pN}}\,r^{pN}_5
\label{40}
 \eeq
 Here $\rho_{pN}$ is the ratio of real to imaginary parts of the forward 
elastic $pN$ amplitudes. We keep $t$-dependence of $\rho_{pA}(t)$ since 
it might be quite steep within the CNI range of $t$.

The CNI effects for nuclei are substantially modified by nuclear 
formfactors \cite{mine,kt} which are steep functions of $t$
 \beqn 
&& \frac{16\,\pi}{(\sigma^{pA}_{tot})^2}\,
\frac{d\,\sigma_{pA}}{d\,t}\,A^{pA}_N(t) 
= \frac{\sqrt{-t}}{m_N}\,
F_A^h(t)\,\biggl\{F_A^{em}(t)\,\frac{t_c}{t}\,
\Bigl[(\mu_p-1)[1-\delta_{pA}(t)\,\rho_{pA}(t)] \nonumber\\ 
&-& 2\,[{\rm
Im}\,r^{pA}_5(t)-\delta_{pA}(t)\,{\rm Re}\,r^{pA}_5(t)]\Bigr] -
2\,F_A^h(t)[{\rm Re}\,r^{pA}_5(t) - \rho_{pA}(t)\, {\rm
Im}\,r^{pA}_5(t)]\biggr\}\ , \label{50}
 \eeqn 
 where
 \beqn \frac{16\,\pi}{(\sigma^{pA}_{tot})^2}\,
\frac{d\,\sigma_{pA}}{d\,t} &=&
\left(\frac{t_c}{t}\right)^2\,\Bigl[F_A^{em}(t)\Bigr]^2 -
2\,[\rho_{pA}(t)+\delta_{pA}(t)]\,\frac{t_c}{t}\,
F_A^h(t)\,F_A^{em}(t)\,\nonumber\\ &+&
\left[1+\rho_{pA}^2(t)-
\frac{t}{m_p^2}\,|r^{pA}_5(t)|^2\right]\,
\Bigl[F_A^h(t)\Bigr]^2\ . \label{60}
 \eeqn
 Here $t_c=-8\pi\alpha/\sigma^{pA}_{tot}$; $\delta_{pA}(t)$ is the 
Coulomb phase for $pA$ scattering calculated in \cite{ktar} with high 
accuracy; the ratio of real to imaginary parts of the $pA$ amplitude,
$\rho_{pA}(t)$ and nuclear formfactors, electromagnetic $F_A^{em}(t)$ 
and hadronic $F_A^h(t)$, are calculated in \cite{ktar} with a realistic
nuclear density. 

\begin{figure}
\includegraphics[height=.3\textheight]{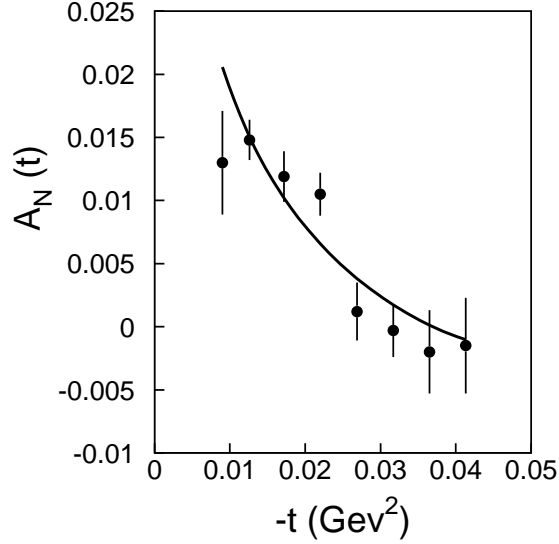}
\caption{The data from \cite{e950} with statistical and systematic 
errors summed quadratically, while the uncertainty in the beam 
polarization is treated and an overall normalization.} 
\label{bnl}
 \end{figure}

A first time precise measurement of $A_N^{pA}(t)$ performed by the E950
collaboration for proton-carbon elastic scattering with $22\GeV$
polarized beam at AGS \cite{e950}. The authors fitted the data with
expressions (\ref{50})-(\ref{60}) and found
 \beq
\Im r_5 = - 0.161 \pm 0.226;\ \ \ \ 
\Re r_5 = 0.088 \pm 0.058\ .
\label{70}
 \eeq
 The authors added linearly the errors, statistical one and two
systematic related to the row asymmetry and the beam polarization. The
resulting error seems to be overestimated and may be treated as an upper
bound. I have repeated the fit adding quadratically the first two errors,
but treating the error in the beam polarization as an overall
normalization. I have arrived to similar central values of the
parameters, but smaller errors which might be treated as a lower bound,
 \beq
\Im r_5 = - 0.156 \pm 0.170;\ \ \ \ 
\Re r_5 = 0.084 \pm 0.042\ .
\label{80}
 \eeq
 The renormalization factor for the beam polarization was found to be 
$N=1.001 \pm 0.120$. The result of the fit and fitted data are depicted 
in fig.~\ref{bnl}

Note that this values of $r_5$ correspond to the iso-scalar part of
elastic $pp$ amplitude.  As far as it is known (with a considerable
uncertainty) at energy $22\GeV$ one may consider using it for polarimetry
at higher energies. This would be appropriate if energy variation of the
Pomeron part of $r_5$ is small and if the sub-leading iso-scalar Reggeon
($\omega$ and $f$) contribution to $r_5$ is small at $22\GeV$. The latter
assumption has been questioned recently and possible corrections for
polarimetry are discussed in \cite{larry}.

Assuming no energy dependence of $r_5$ one can use (\ref{70}) to predict
$A_N^{pp}$ at energy $200\GeV$ and compare with the E704 data (including
data at larger $t$, see \cite{e704}), as is depicted in
Fig.~\ref{pp-pomm} (right panel)  by thick solid curve, while the
corridor related to the errors in (\ref{70}) is shown by thin solid
curves. In spite of large uncertainties in (\ref{70}) one may conclude
that data do not support such a prediction. Of course this comparison is
based of unjustified assumption of no energy dependence of $r_5$,
nevertheless, the observed disagreement should be considered as a
warning.

\section{Regge factorization: analysis of $\pi N$ data}

\begin{figure}
\includegraphics[height=.3\textheight]{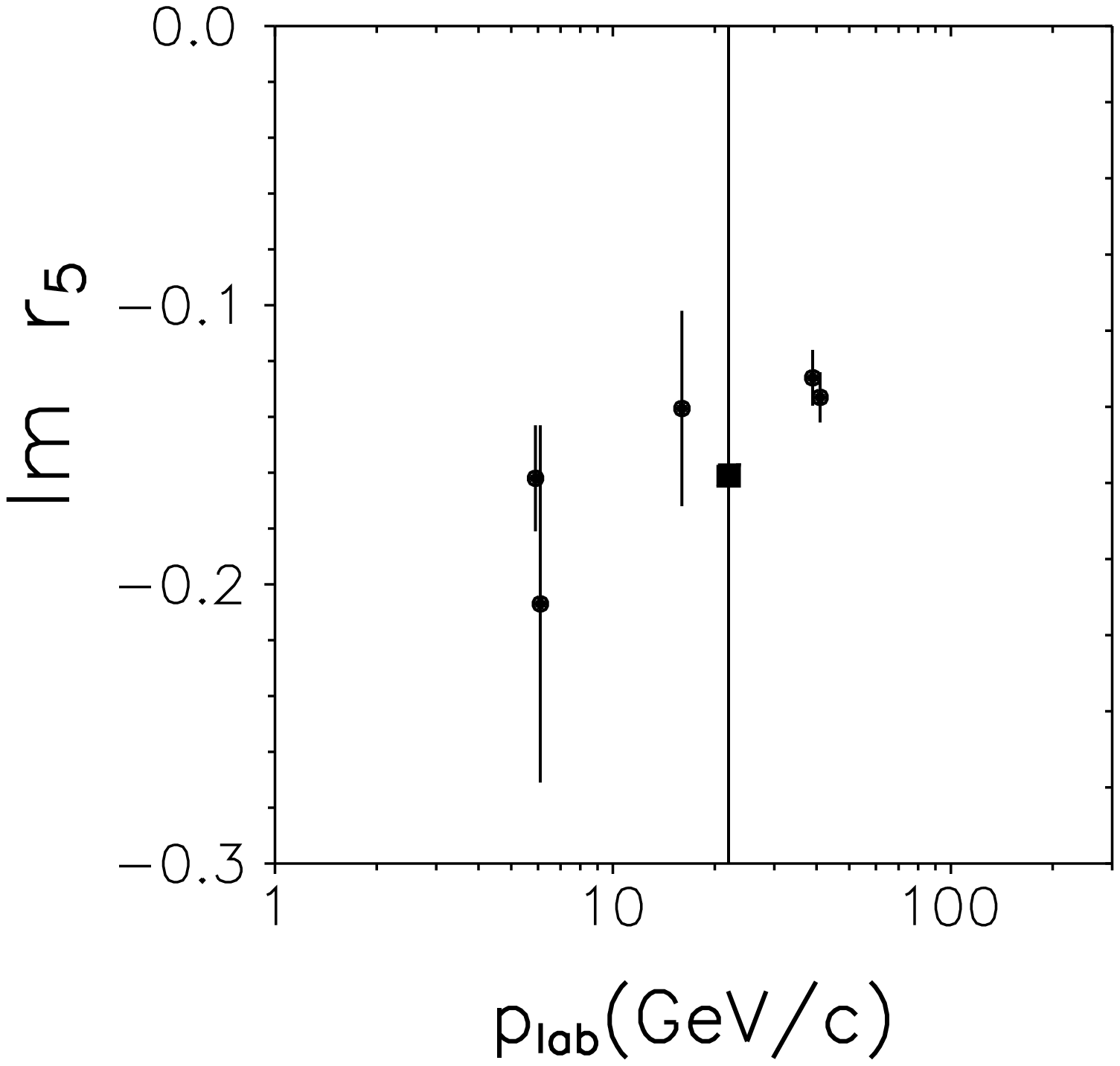}
\includegraphics[height=.3\textheight]{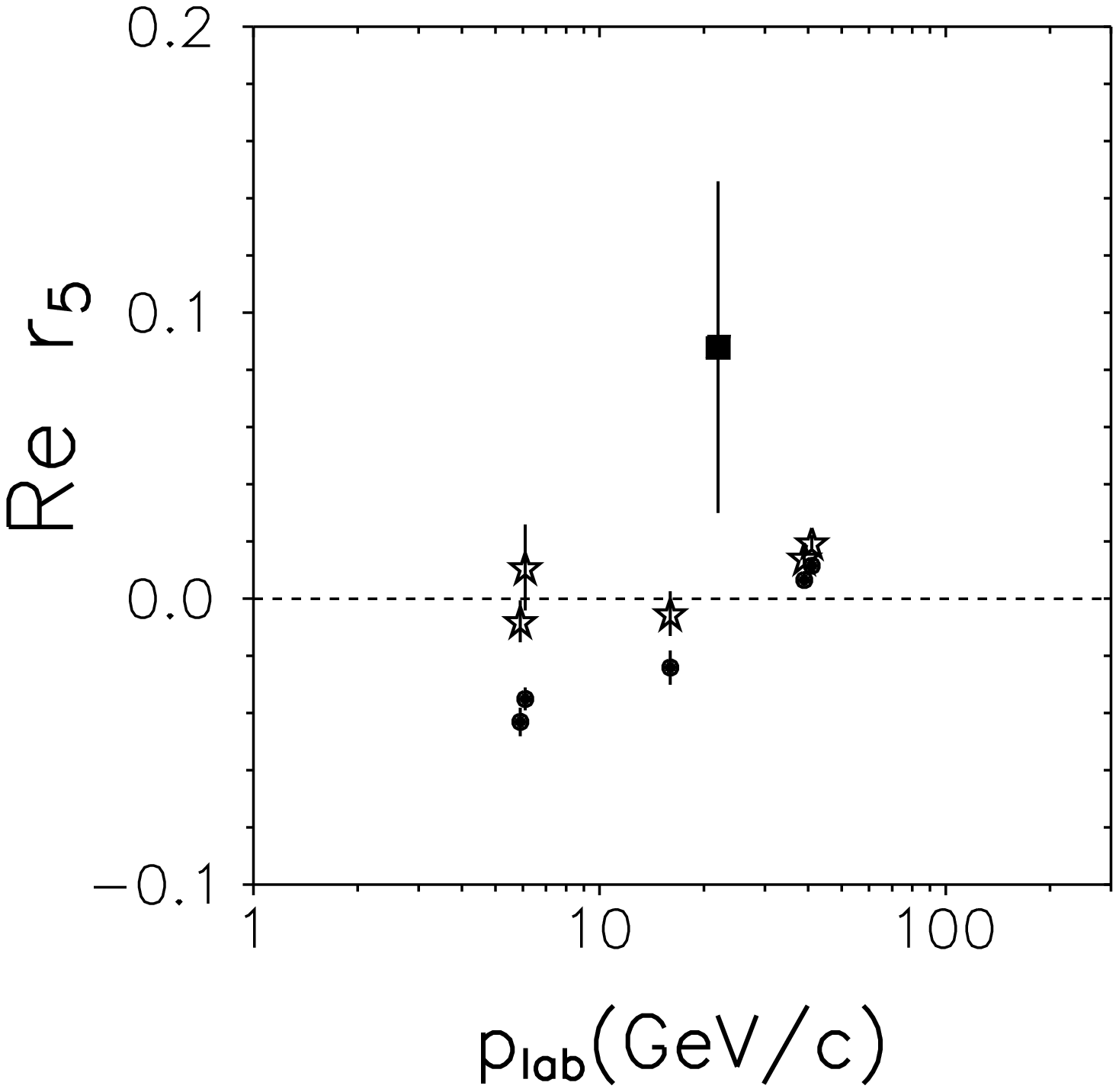}
 \caption{Comparison of the results of the E950 experiment at BNL (square 
points) with the results of amplitude analyses \cite{amplitudes} of $\pi 
N$ data. Left panel: data for $\Im r_5$ are shown by full 
round dots. Right panel: round dots show the phase uncorrected results of 
\cite{amplitudes} for $\Re r_5$, star points are corrected for the phase 
of the non-flip amplitude.}
\label{piN}
 \end{figure}

The iso-scalar part of $r^{NN}_5$ extracted from $pA$ may be compared
with $\pi N$ data. They should be related provided Regge factorization
holds. Amplitude analyses of $\pi N$ elastic and charge exchange
scattering up to energy $40\GeV$ are available \cite{amplitudes} and the
results contain $r_5(t)$ for iso-scalar part of the scattering amplitude.
It turns out that all the analyses demonstrate no $t$-dependence of
$r_5(t)$ within error bars for $|t| < 0.5\GeV^2$, what is not surprising
since the $\sqrt{-t}$ factor is removed. In order to reduce uncertainties
data for iso-scalar $r_5(t)$ for each analysis was fitted by a constant
within this $t$-interval. The results for $\Im r_5$ are depicted in
Fig.~\ref{piN} by round points, while the E950 value is shown by a
square.

Apparently, the $\pi N$ data prefer negative $\Im r_5$, however they do
not specify energy dependence. Within large error bars they are
consistent either with no energy dependence, or with $\Im r_5$ rising
with energy. The former case would correspond to a net contribution of
the Pomeron spin-flip, while the latter possibility would mean that
$f$-Reggeon contribution to $\Im r_5$ exists and is negative. Thus, we
conclude that $\Im r_5^f\leq 0$. Since the phase of the $f$-amplitude is
given by the signature factor, $\eta(t) = i - \cot[\pi\alpha_f(t)/2)]$ we
should expect from this consideration that $\Re r_5^f\geq 0$.

Data for $\Re r_5$ extracted from the same analyses \cite{amplitudes} are
depicted by round points in Fig.~\ref{piN} (right panel). The real part
of the spin flip amplitude was determined in those analyses relative to
the imaginary part of the non-flip amplitude, i.e. assuming it pure
imaginary. Thus, one should introduce a correction for a nonzero real
part of the non-flip amplitude, $\Delta \Re F_{+-} = \rho_{\pi N}\,
\Im F_{+-}$. Using $\Im r_5$ found above, new corrected values for $\Re
r_5$ were determined and plotted in Fig.~\ref{piN} (right panel) by star
points.  These results are in agreement with the above expectation $\Re
r_5^f\geq 0$, preferring, however, zero and energy independent value. The
point from the E950 experiment shown by a square is somewhat higher, but
still is compatible with these results.

Thus, available amplitude analyses of $\pi N$ data at energies $6 -
40\GeV$ indicate at the dominance of the Pomeron amplitude with
 \beq
\Im r_5 \approx -0.12;\ \ \ \ \Re r_5 \approx 0\ ,
\label{90}
 \eeq
 and vanishing contribution of the $f$-Reggeon.

\section{Comparison with theoretical expectations}

One can find in the literature a variety model predictions for the 
spin-flip part 
of the Pomeron amplitude. Many of them are collected 
and discuss in \cite{review}. Here we list them briefly mentioning the 
underlying physical ideas.
\begin{itemize}
 \item
 Treating the gluon-quark vertex as an analog to the iso-scalar
photon-proton one can relate the anomalous color-magnetic moment of a
quark to the iso-scalar part of the anomalous magnetic momentum of the
proton \cite{mr}. After installation of such a quark-gluon vertex into
the two-gluon model for the Pomeron one gets \cite{mr}, \underline{$\Im
r_5 = 0.13$}. Although the order of magnitude is correct, the sign is
opposite to data presented in Fig.~\ref{piN}.
 \item
 Helicity of the proton is not equal to the sum of quark helicities.
Therefore, the proton may flip its helicity even if quarks do not (as the
leading order pQCD predicts). A quark-diquark model of the proton leads
to nonzero \underline{$\Im r_5 = - (0.05 - 0.15)$}, dependent on the 
diquark size ($0.5 - 0.2\fm$) \cite{kz}. Within the uncertainty this 
prediction agrees with the data.
 \item 
 Modeling the Pomeron-proton coupling via two pion exchange
\cite{pumplin,boreskov} one arrives at a conclusion that iso-scalar
Reggeons ($\Pom,\ f,\ \omega$) are predominantly spin non-flip, while
iso-vectors ($\rho,\ a_2$) mostly flip the proton spin. Prediction of
\cite{boreskov} for the Pomeron is \underline{$\Im r_5 = 0.06$}, what has
incorrect sign.  A similar pion cloud model developed \cite{goloskokov}
with some differences in details predicts \underline{$\Im r_5 = - 0.3,\
\Re r_5 = -0.06$}, what also disagree with the data.
 \item 
 The phenomenological model \cite{soffer} assuming that the spin-flip
part can be deduced from the impact parameter distribution of matter in
the proton and fitted to data predicts correct sign, \underline{$\Im r_5
\approx -(0.01-0.02)$}, but modulo too small value.
\end{itemize}

 One should not treat this comparison as a way to confirm or reject
models. None of the models under discussion may pretend to be a dominant
mechanism. The dynamics suggested by other models can contribute as well.

Note that analysis of $pp$ elastic data performed in \cite{enrico} led to
parameters \underline{$\Im r_5 = -0.054$} which is too small, but has the
right order of magnitude and correct sign compared to data plotted in
Fig.~\ref{piN}.  The analysis performed in \cite{enrico} was based on a
specific modeling of the odderon amplitude which introduces a strong
sensitivity of polarization to $r_5$. Besides, the contribution of the
sub-leading Reggeons largely contributing to $r_5$ ($\rho$, $a_2$) was
neglected, instead this this contribution was attributed to the Pomeron.

\section{$\pi N$ vs E950 data: how shaky is the theory bridge?}

The results of amplitude analyses of $\pi N$ data are good news for 
polarimetry at RHIC. Absence of energy dependent contribution of 
iso-scalar sub-leading Reggeons to $r_5$ suggested by the data
would allow one to use the result of measurement of $A_N$ by the E950 
experiment for polarimetry at higher RHIC energies. However, the central
value of $\Re r_5$ which follows from the E950 data is different from 
zero and indicates that the Reggeon contribution might be important.
Then, one may expect $r_5$ to vary with energy and the polarimetry gets 
an uncertainty.

Moreover, the fitting parameters $\Re r_5$ and $\Im r_5$ strongly 
correlate as it is demonstrated in \cite{e950}. For example, if to 
enforce and fix $\Re r_5=0$, the $\chi^2$ doubles and $\Im r_5$
changes sign. Thus, it is difficult to bring together the results of 
study of different reactions $\pi N$ and $pC$.

Facing such a problem one should check how reliable are assumptions done
in order to make a link between iso-scalar amplitudes in $pp$ and $\pi N$
scattering. 
 \begin{enumerate}
 \item First of all, how precise is factorization connecting $r_5$ in
$pp$ and $\pi N$? In all the models listed above it is provided. Even in
the two-gluon model which does not obey Regge factorization, $r_5$ must
be the same for $\pi N$ and $pp$. It is known, however that that Regge
cuts corresponding to eikonal multi-Pomeron rescatterings violate Regge
factorization. However, as is discussed above and proven in \cite{bethe}
these corrections do not alterate $r_5$.
 \item
 Only sub-leading Reggeon, $f$, contributes to the iso-scalar amplitude in
$\pi N$ scattering, while both $f$ and $\omega$ are present in $pp$.
Moreover, in order to respect duality $f$ and $\omega$ should be exchange
degenerate, i.e. their contributions are expected to add up in $\Re r_5$
and nearly cancel in $\Im r_5$. This is different from $\pi N$ where
$f$-Reggeon should contribute equally to $\Re r_5$ and $\Im r_5$.
However, this difference does not explain the observed difference between
$\pi N$ and $pp$. If $f$-Reggeon does not contribute to $r_5$ in $\pi N$,
according to factorization and exchange degeneracy both the $f$ and
$\omega$ contributions to $pp$ must be zero as well.
 \end{enumerate}

Although we did not find any good reason to disbelieve the theoretical
link between $\pi N$ and $pp$, it is still possible that this is the
origin of the problem. On the other hand the observed contradiction is
not dramatic since the errors of the E950 data are pretty large. In order
to progress further, the accuracy of $A_N$ in $pC$ elastic scattering
should be improved.

\section{Direct information from NN data}

\begin{figure}
\includegraphics[height=.3\textheight]{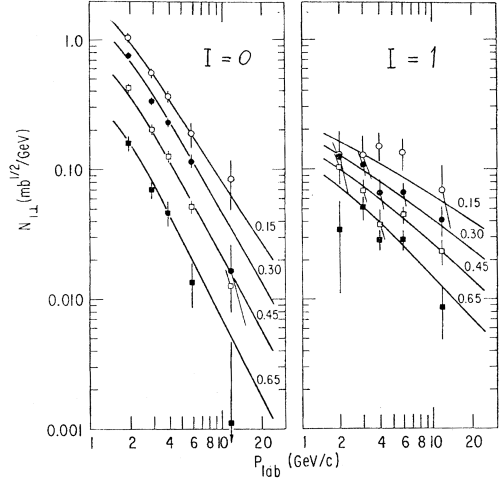}
\caption{Dependence of the spin-flip amplitude on lab momentum
for iso-scalar (left panel) and iso-vector (right panel) exchanges.
Points are the result of the analysis of data on elastic and 
charge-exchange $pp$ and $pn$ scattering performed in \cite{kramer} for 
different bins in $t$, and the curves are the results of Regge fit.}
 \label{NN}
 \end{figure}

 There is another narrow place in the theoretical bridge between $\pi N$
and $NN$ reactions: it might be a contribution to $NN$ of sub-leading
Reggeons which are forbidden for $\pi N$. For instance, besides $\omega$
there might be other iso-scalar mesons which are suppressed or forbidden
(e.g. have negative $G$-parity) for $\pi n$ scattering.  This was
suggested in \cite{berger} as $\epsilon(0^{++})$ and $\omega'(1^{--})$
exchange degenerate Reggeons. Indeed, analysis \cite{kramer} of data for
$pp$ and $np$ elastic scattering up to $12\GeV$ shown in Fig.~\ref{NN}
demonstrates an iso-scalar spin-flip $NN$ amplitude (left panel) which
falls with energy much steeper than iso-vector one (right panel).
 The iso-scalar Regge trajectory turns out to be displaced by one unit
down compared to the $\rho$-Reggeon trajectory: $\alpha_\epsilon(t) =
\alpha_\rho(t) - 1 = -0.5 + 0.9t$ \cite{kramer}.

It is important to establish whether the large value of $\Re r_5$
observed by the E950 experiment is related to the tail of this low-energy
mechanism. If so, then $\Re r_5$ will steeply vanish at higher energies
what should affect the polarimetry. In this case the shape of $A_N(t)$ in
the CNI region would change substantially (not supported by preliminary
data at $100\GeV$).

One can estimate such a low-energy contribution to $r_5$ at $20\GeV$
relying on the extrapolation of the iso-scalar spin-flip $NN$ amplitude
done in \cite{kramer} depicted in Fig.~\ref{NN}.
 The iso-scalar amplitude is determined by measurement of $A_N$ and cross
sections of elastic and charge-exchange $pp$ and $pn$ scattering,
 \beq
N^0_{1\perp}=\left[(A_N\sigma)_{pp}+
(A_N\sigma)_{pn} -{1\over2}(A_N\sigma)_{cex}\right]
\Bigl/(4\,|N^0_0|)\ ,
\label{110}
 \eeq
 where $|N^0_0|^2=(\sigma_{pp}+\sigma_{pn})/2$. At $t=-0.15$ this
amplitude is predicted to be, $N^0_1 \approx 0.03\,\sqrt{\mb}/\GeV$. The
non-flip amplitude equals to $N^0_0 \approx
\sigma_{tot}/(4\sqrt{\pi})\exp(5t)\approx 4.2\,\sqrt{\mb}/\GeV$. Taking
into account the factor $\sqrt{-t}/m_N$ in $N^0_1$, one arrives at the
estimate at $t=-0.15\GeV^2$,
 \beq
\Re r_5(p_{lab}=22\GeV/c) \approx 0.02\ ,
\label{100}
 \eeq
 which is too small to explain the value of $\Re r_5$ in (\ref{70}).

This estimate agrees well with the measurements of single-spin asymmetry
in $pp$ and $pn$ performed at $24\GeV$ at BNL \cite{pp}. Neglecting the
small charge-exchange contribution (it steeply falls with energy) in
(\ref{110}) one gets at $t=-0.15\GeV^2$,
 \beq
\Re r_5(p_{lab}=24\GeV/c) =  0.016 \pm 0.010\ ,
\label{120}
 \eeq

 Thus, both extrapolation of Argonne data to higher energies and
direct measurements at AGS at $24\GeV$ confirm that $\Re r_5$ is about 
order of magnitude smaller than what follows from the E950 data. 

It is also very improbable that $r_5(t)$ could vary substantially at $0 <
-t < 0.15\GeV^2$. As it was mentioned above, in $\pi N$ data $r_5$
remains unchanged up to $-t=0.5\GeV^2$.

\section{conclusions and outlook}

The E950 experiment has provided first high statistics measurements for
CNI asymmetry in proton-carbon elastic scattering. On the one hand, these
data bring information about the spin-flip part of the hadronic amplitude
which is tempting to associate with the Pomeron. On the other hand, if it
true, one can use the found parameters for $r_5$ to predict $A_N(t)$ at
higher energies and use $pC$ scattering as a polarimeter at RHIC.

At the same time, amplitude analysis of data for $\pi N$ and $NN$ elastic
and charge-exchange scattering allow to single out the iso-scalar part of
the spin-flip amplitude. The values of $\Re r_5$ extracted from these
data are sufficiently small to be neglected. This is a great news for the
CNI polarimetry which can be safely used at high energies. This value of
$\Re r_5$ is, however, much smaller than found from the E950 data. To
resolve this controversy one needs new and more precise data for CNI spin
asymmetry and in a wider energy range. 

 \begin{theacknowledgments}
 I am thankful to Larry Trueman for interesting discussions and to Yousef
Makdisi for inviting me to speak at the Conference. This research was
performed during visit at the BNL Nuclear Theory Group, and I am grateful
to Larry McLerran for hospitality. This work has been supported by a
grant from the Gesellschaft f\"ur Schwerionenforschung Darmstadt (GSI),
grant No.~GSI-OR-SCH, and also by the grant INTAS-97-OPEN-31696.
 \end{theacknowledgments}

\end{document}